\documentclass[superscriptaddress]{revtex4}
\usepackage{epsfig}
\usepackage{graphicx}
\usepackage{color}

\begin{document}

\title{Hybrid improper ferroelectricity in a
 Multiferroic\\ and Magnetoelectric 		
hybrid organic-inorganic perosvkite}

\author{Alessandro Stroppa}
\affiliation{CNR-SPIN, L'Aquila, Italy}

\author{Paolo  Barone} 
\affiliation{CNR-SPIN, L'Aquila, Italy}

\author{Prashant Jain}
\affiliation{Los Alamos National Lab,  30 Bikini Atoll Rd  Los Alamos, NM 87545-0001 (505) 664-5265}

\author{Jean Manuel  Perez-Mato}
\affiliation{Departamento de Fisica de la Materia Condensada,
Facultad de Ciencia y Tecnologia, UPV/EHU, Bilbao (Spain)}

\author{Silvia Picozzi}
 \affiliation{CNR-SPIN, L'Aquila, Italy}

\begin{abstract}
 There is great interest in hybrid organic-inorganic materials such 
as metal-organic
frameworks (MOFs). The   compounds  
 [C(NH$_{2}$)$_{3}$]M(HCOO)$_{3}$, where M=Cu$^{2+}$ or Cr$^{2+}$ are Jahn-Teller (JT) active ions, 
are MOF with perovskite topology which crystallizes in polar space group 
Pna2$_{1}$. In inorganic compounds, octahedral 
tilting and Jahn-Teller structural distortions are usually
non-polar distortions. However, in this MOF cooperative interactions
 between the
antiferro-distortive distortions of the framework and the C(NH2)$_{3}$
 organic cation
via hydrogen bonding breaks the inversion symmetry and induces a ferroelectric
polarization.[Angew. Chem. Int. Ed. \textbf{50}, 5847, 2011]
 Our ab-initio study supports the picture of an orbital-order-induced
ferroelectricity, a rare example of dipolar ordering caused by
 electronic degrees of
freedom. Moreover,  the switching of
 polarization direction implies
the reversal of the weak ferromagnetic component. 
The microscopic mechanism of the multiferroicity and magnetoelectric coupling 
in this  JT-based MOF with ABX$_{3}$ perovskite structure displays  a
Hybrid Improper Ferroelectric (HIF) state, arising from a trilinear coupling between
different structural  deformations that comprise tilting, rotations and Jahn-Teller
distortions of both  the BX$_{3}$ framework and the organic cation at the A sites. 
 Since these distortion modes in perovskite-inorganic compounds usually freeze-in at elevated
temperatures, the trilinear coupling in MOF compounds may provide an interesting
route towards high-temperature multiferroicity.
 These results therefore offer
an important starting point for tailoring multiferroic properties
 in this emerging
class of materials for various technological applications. In particular, 
the high tunability  of the  ferroelectric polarization by means of the modification of the organic A cation has been recently shown[J. Am. Chem. Soc. \textbf{135}  18126 (2013)]
 \end{abstract}
\maketitle
 
Email: alessandro.stroppa@spin.cnr.it
\\
\\

Metal-organic frameworks (MOFs) are  hybrid crystalline
compounds comprised of extended ordered networks formed from organic 
linkers and metal cations, often forming porous materials at the interface
 of molecular coordination chemistry and materials science. 
They show unique properties arising from  organic-inorganic 
duality\cite{Yaghi} which has already resulted in an unprecedented 
variety of physical properties in a single class of materials  and
applications, such as gas storage, exchange or separation,
catalysis, drug delivery, optics, and magnetism.\cite{Ferey,Ferey1} 
An additional feature is the possibility  
of creating ideally infinite new MOFs by 
varying  inorganic/organic components,  molecular
topologies,  organic linkers, etc.\cite{Highlights,Highlights1,Highlights2,Highlights3,PRLMOF,Baburin,
Tsukerblat} 
All these degree of freedoms can be exploited for 
a rational design of new materials with enhanced functionalities.

MOFs with ABX$_{3}$ perovskite 
structure  or   closely related 
superstructures of this chemical \textit{chameleon} have attracted 
much attention since they show promising properties in
 areas that have traditionally
been dominated by inorganic materials, for example
  magnetism and
 ferroelectricity.\cite{NewRoutes,NewRoutes1,NewRoutes2,NewRoutes3} 
Combination of spontaneous magnetic and ferroelectric order 
in a single material, \textit{i.e.} multiferroicity (MF),
 is of great technological and fundamental importance, in particular 
when both orders are coupled through a sizeable magneto-electric 
coupling (ME). Despite the large activity devoted to 
multiferroics,\cite{Spaldin} most of the past and current 
studies have been focused on inorganic compounds, though  mainly in the 
family of perovskite-like oxides. Among these, a strong ME coupling is 
expected in
magnetically driven improper ferroelectrics, such as rare-earth manganites 
or delafossite oxides,
where the magnetic ordering is responsible for  spontaneous
electric polarization.\cite{cheongmostovoy} Unfortunately,
 the  symmetry-breaking is usually associated with frustrated
magnetism   displaying antiferromagnetic or spiral order, and both 
 measured electric polarization and critical
temperatures are usually too small for any device applications.\cite{cheongmostovoy} The large ferroelectric polarization of
proper multiferroics such as BiFeO$_3$, on the other hand, originates from a polar lattice
instability that is generally not coupled to any magnetic instability leading to a net 
magnetization.\cite{BFO} 

Very recently, a third class of 
magnetoelectric multiferroics has been suggested, where a lattice instability is responsible for both ferroelectricity and the appearance of a weak-ferromagnetic (WFM)
ordering, thus allowing for a potentially  large ME coupling. The
key ingredient is a trilinear coupling between a (stable) polar mode and two nonpolar
instabilities, usually octahedron tilting and rotations in   layered or double  perovskites, as recently found in
some inorganic compounds,  where layered cation ordering  leads to 
the required symmetry breaking.\cite{rot1,rot2,rot3,rot4,rot5} 
 The mechanism has been called 
 ``hybrid improper ferroelectricity'', 
alluding to the fact that polarization appears
as a secondary order parameter from the trilinear coupling 
to non-polar unstable modes.\cite{rot2} Remarkably, the experimental 
evidence of the proposed mechanism in a class of inorganic compounds 
has been recently put forward.\cite{rotexp}

Here, we theoretically predict that a Cr based MOF (hereafter called Cr-MOF) of ABX$_{3}$ perovskite topology [C(NH$_{2}$)$_{3}$]M[(HCOO)$_{3}$], not yet synthesized,\cite{KeLi}
should be a new multiferroic  and we predict a
\textit{very large} WFM component which 
is coupled to ferroelectricity. We will show that it belongs 
to the class of ``hybrid improper'' multiferroics.\cite{rot2}
Specifically,
we found that the rotational mode associated to A-site molecules  
and the Jahn-Teller (JT) distortions 
of the Cr$^{2+}$ ions  
are unstable primary modes 
while the polar mode is soft but stable. The ferroelectricity arises only because the stable polar mode is coupled
to the two nonpolar unstable modes, through a trilinear coupling. 
  Microscopically, this coupling is
expected to be mediated by hydrogen bondings connecting the guanidinium cations with the distorted metal framework, as
discussed in Ref.\cite{Stroppa}. Here, for the first time, the ``hybrid improper'' nature of 
ferroelectricity 
involving JT modes is found and discussed in a MOF with ABX$_3$ structure.
Furthermore, we propose a microscopic model showing that the WFM  originates from the peculiar
magnetic anisotropy  of  JT ions in combination  with the strongly distorted structure of
 the metal framework.\cite{moriya2}
 Our theoretical predictions 
point therefore to a central role of Jahn-Teller interaction in driving both
FE and ME properties. 
Finally, we predict by means of Monte Carlo calculations a critical
 temperature $T_c\sim 40$ K
for the WFM phase, one of the
 largest among the class of the ABX$_{3}$ MOF compounds.\cite{Tc1,Tc2}

A new family of MOFs in perovskite ABX$_{3}$ topology has been synthesized,\cite{KeLi} 
such as [C(NH$_{2}$)$_{3}$]M[(HCOO)$_{3}$], where A=C(NH$_{2}$)$_{3}$$^{+}$ is the  guanidinium cation, M$^{+2}$ is a divalent metal atom (Mn, Fe, Co, Ni, Cu or Zn) and X is the formate HCOO$^{-}$.
The Cu based MOF, which is the only 
one of the series containing a JT ion,
 crystallizes in a polar space group, Pna2$_{1}$, displaying both ferroelectricity and WFM, while all the others are 
isostructural and  crystallize in a non-polar Pnna space group.\cite{KeLi,Stroppa} 

\begin{figure}[h]
\centering
\includegraphics[width=0.8\textwidth,angle=0,clip=true]{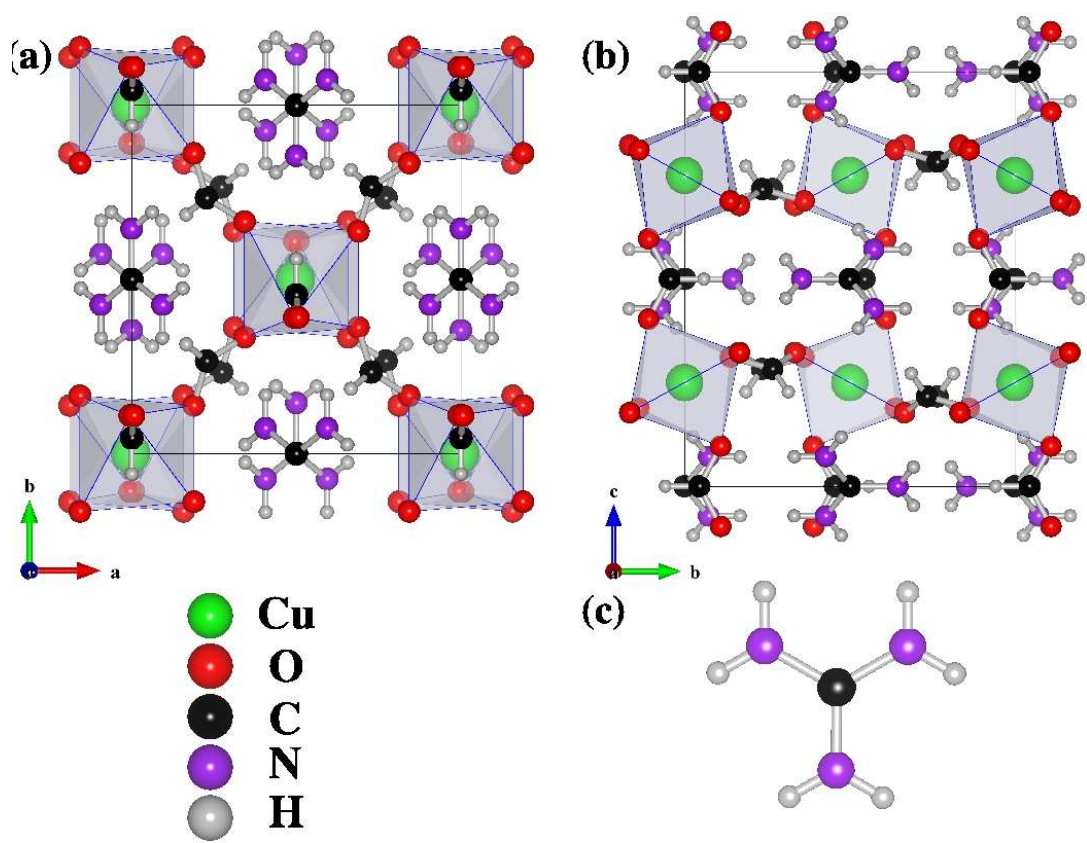}
\caption{ 
(a) top and (b) side view of the crystalline unit cell; 
(c) Guanidinium cation.}\label{Fig1}
\end{figure}

Here we combined density functional theory calculation, symmetry analysis and model Hamiltonian in order to investigate the  case of the JT 
active Cr$^{+2}$ ion.  
Although it may be difficult to stabilize the Cr ion in a
 low oxidation state $^{+2}$, very 
recently the homologous inorganic compound K$^{+1}$Cr$^{+2}$F$^{-1}$$_{3}$ has been synthesized and studied over a wide range of temperature.\cite{Margadonna}
The top and side view of the Cr-MOF  are shown respectively in Fig.\ \ref{Fig1} (a) and (b), while in Fig.\ \ref{Fig1} (c) 
 the  A-group guanidinium is shown.
Starting from the crystal structure of the Cu-MOF, we 
replaced   the Cu$^{+2}$ ($t_{2g}^{6}e_{g}^{3}$, twofold hole degeneracy)   ions 
by Cr$^{+2}$ active ions ($t_{2g}^{3}e_{g}^{1}$, twofold electron degeneracy).
In both cases, they are JT active. For Cr$^{+2}$, as a result of strong
 Hund's rule coupling,  spin of the $e_{g}$ electron is parallel to 
spins of the $t_{2g}$ electrons on   same site. 
The relaxed polar crystal structure $Pna2_{1}$  turns out to be  more stable than the centric $Pnna$ space 
group by about $\sim$ 0.08 eV/formula unit. The CrX$_{6}$  octahedra (X=HCOO)
 are strongly axially distorted, comprising \textit{medium} Cr-X bonds 
 $\sim$ 2.033\ \AA\ along   $c$ axis, and alternating \textit{long} 
2.358\ \AA\ and \textit{short} 2.010\ \AA\ in  $ab$ plane. 
Orientation of  long  Cr-X bonds in   $ab$ plane is 
a direct result of the cooperative Jahn-Teller
distortions (CJTDs) resulting in antiferrodistortive pattern 
of long and short bonds in   $ab$ plane:    long  bonds form 
two ferrodistortive sublattices  characterized by a parallel orientation, 
however  the orientation 
is rotated by $\sim$ 90$^{\circ}$ from one sublattice to another.
Such an ordering of 
long bonds is compatible with an antiferrodistortive 
ordering of the 3d$_{3x^2-r^2}$ and 3d$_{3y^2-r^2}$ orbitals (OO) 
in   same plane. The orbital pattern
is  the same when moving along the $c$ axis. 
Each octahedron is tilted  with respect to the $c$ axis by 
$\sim$ 31$^{\circ}$.

\begin{figure}
\centering
\includegraphics[width=0.8\textwidth,angle=0,clip=true]{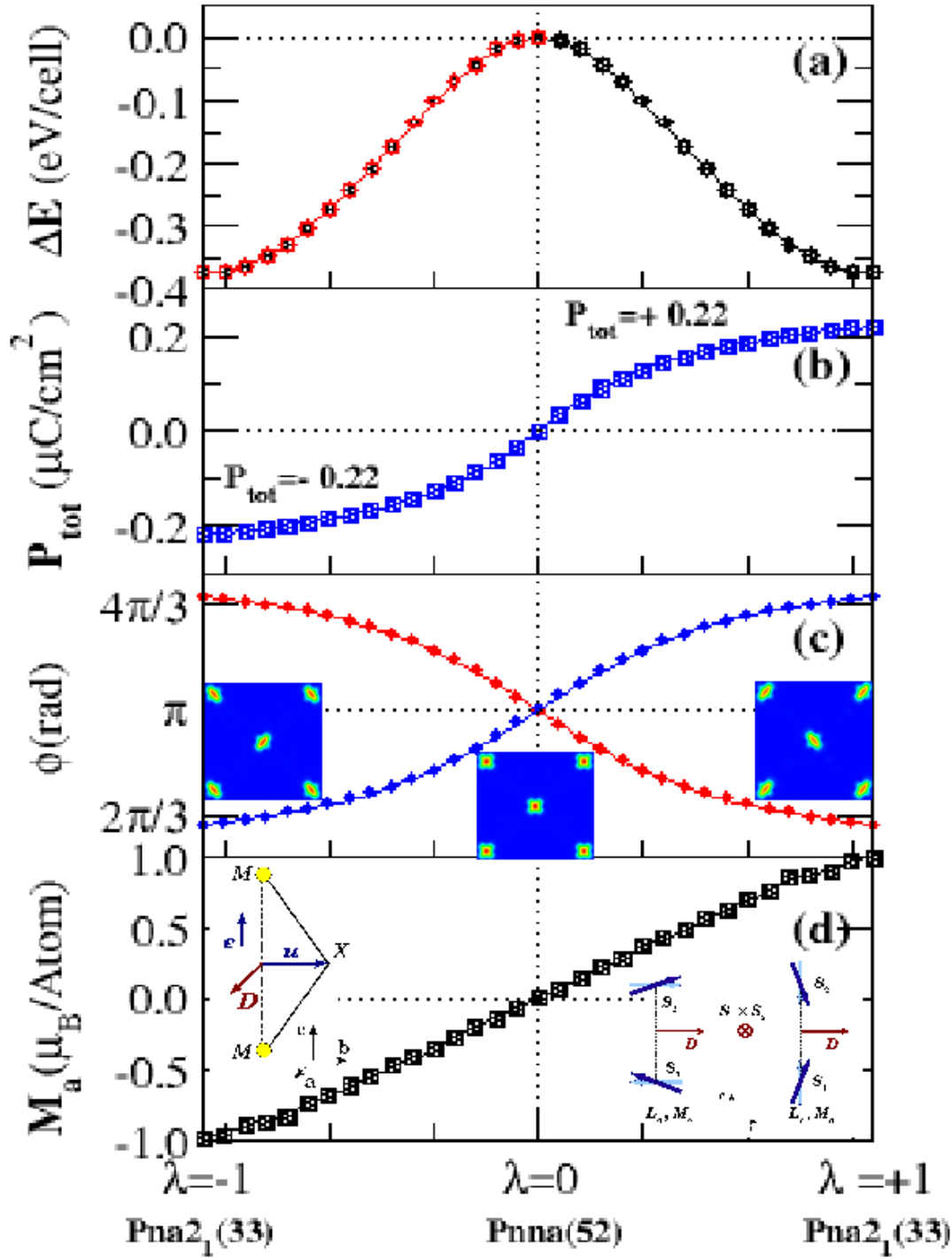}
\caption{ (a) Variation of the total
 energy as a function of the structural  distortion from 
the centric to the polar structures (switching path); 
(b) Ferroelectric polarization,  
(c) variation of the orbital ordering of the
$e_{g}$ electron and the JT $\phi$ phase, (d) variation of the WFM component along the switching path;  
in (d), inset from the left: schematic view of the Dzyaloshinskii vector,
defined as ${\bf D}_c\propto {\bf u}\times
{\bf e}$, where ${\bf u}\parallel b$ is the displacement vector of $X$ ligand  and ${\bf e}\parallel c$ is the unit vector
connecting neighboring spins belonging to different layers; inset from the right, 
 symmetry-allowed WFM configurations 
shown, showing that spin chirality is 
always perpendicular to Dzyaloshinskii vector $\bf D$,
resulting in no Dzyaloshinskii-Moriya interaction.}\label{Fig2}
\end{figure}

In Fig.\ \ref{Fig2} (a) we report the  variation of
total energy as a function of
normalized amplitude $\lambda$ of the structural distortion 
connecting the centric Pnna phase ($\lambda$=0) to polar one ($\lambda$=$\pm$1). 
The expected double-well profile is characteristic of a 
switchable ferroelectric system.\cite{Lines}
The electric  polarization as a function of $\lambda$ is shown in
 Fig.\ \ref{Fig2} (b), displaying a value 0.22\ $\mu C/cm^{2}$
in the most stable structure ($\lambda$=$\pm$1). 
In Fig.\ \ref{Fig2} (c), we show   changes of the antiferrodistortive
pattern from   centrosymmetric to the polar
phase by introducing the  Jahn-Teller coordinate vector with components
Q$_{2}$ and Q$_{3}$, where $Q_{2}=(l-s)/\sqrt{2}$ and 
 $Q_{3}=(2m-l-s)/\sqrt{6}$;\cite{phi1,phi2} here 
  $l$ and $s$ refer to   lengths of the long and short
bonds, respectively, in  $ab$ planes and $m$ to the lengths of
  CuO$_{ap}$ bonds and the JT phase is defined as $\phi$=tan$^{-1}Q_{2}/Q_{3}$.
Due to the JT effect,  degeneracy of the $e_{g}$ orbitals is lifted and 
the occupied orbital is defined as  $|\theta>=cos\frac{\phi}{2}|3z^{2}-1>+sin\frac{\phi}{2}|x^{2}-y^{2}>$.
One can monitor the antiferro-distortive pattern via   changes in $\phi$.
For  $\lambda=1$ (positive
polarization), $\phi$ is  $\sim 4 \pi/3$ for the
octahedra elongated along [1,1,0]   and  $2 \pi/3$
for octahedra elongated
along [\={1},1,0]. For $\lambda=-1$ (negative
polarization),  the Jahn-Teller phase is swapped between 
the two octahedral sublattices, {\it i.e.} the direction of the
elongated axes are interchanged [see Fig.\ref{Fig2} (c)], while
the antiferro-distortive pattern 
disappears for $\lambda=0$ (centrosymmetric)
 phase, {\em i.e.} $\phi=\pi$, $Q_{2}=0$ and $l=s$.
It is remarkable to note the correlated 
behavior of $\phi$ and that of
the polarization, as a function of $\lambda$. 
This is confirmed by a fit testing procedure using 
$P(\lambda)=A_{0}atan(A_{1} \lambda)$ in Fig.\ref{Fig2} (b), 
which shows a very good  agreement between
the data-points calculated by using 
modern theory of polarization and the data-points calculated with the 
proposed fitting function. Note that 
by definition, the JT $\phi$ phase has the same functional behaviour.
The spin-polarized charge density of the $e_{g}$ electron
 is also shown in the inset of Fig. \ref{Fig2} (c), showing that
the ferroelectric order, the  AFD JT distortions and
the related orbital order are clearly correlated.
It is important to note that 
in standard perovskite ABX$_{3}$ compounds antiferrodistortions 
are never associated to a polar behavior.\cite{bersuker}

The magnetic structure of the Cr-MOF is strongly
influenced by the antiferrodistortive JT distortions. 
The orbital ordering induced by the CJTDs effectively 
confines the magnetic interactions to linear 
chains along the $c$ axis\cite{GKA,towler1995}.
In agreement with Goodenough-Kanamori rules, 
the super-exchange interaction $J_c$ along 
the $c$ axis is antiferromagnetic and 
larger than the ferromagnetic exchange interaction $J_{ab}$ 
in the $ab$ planes, due to the orthogonality of the JT magnetic orbitals; 
as a result, the ground state displays an  
A-type antiferromagnetic (AFM-A) configuration 
(see Table S1 in supplemental material).
From total-energy calculations and assuming   
 $S=2$  spins for  Cr ions, 
 we estimate the exchange couplings 
as $J_c\simeq 0.823~ meV$ and $J_{ab}\simeq -0.452~ meV\,
$, confirming   strong anisotropies of magnetic exchanges.
According to our $ab-initio$ calculations,
 the spins deviate from the collinear arrangement giving rise to a WFM component   along the  crystallographic $c$ 
axis as large as $\sim$ 1 $\mu_{B}$. 
This ferromagnetic component is clearly correlated  with the 
ferroelectric polarization, as shown in Fig. \ref{Fig2} (d), 
\textit{i.e.} the Cr-MOF is expected 
to be a magnetoelectric compound.

\begin{figure}
\centering
\includegraphics[width=0.7\textwidth,angle=0,clip=true]{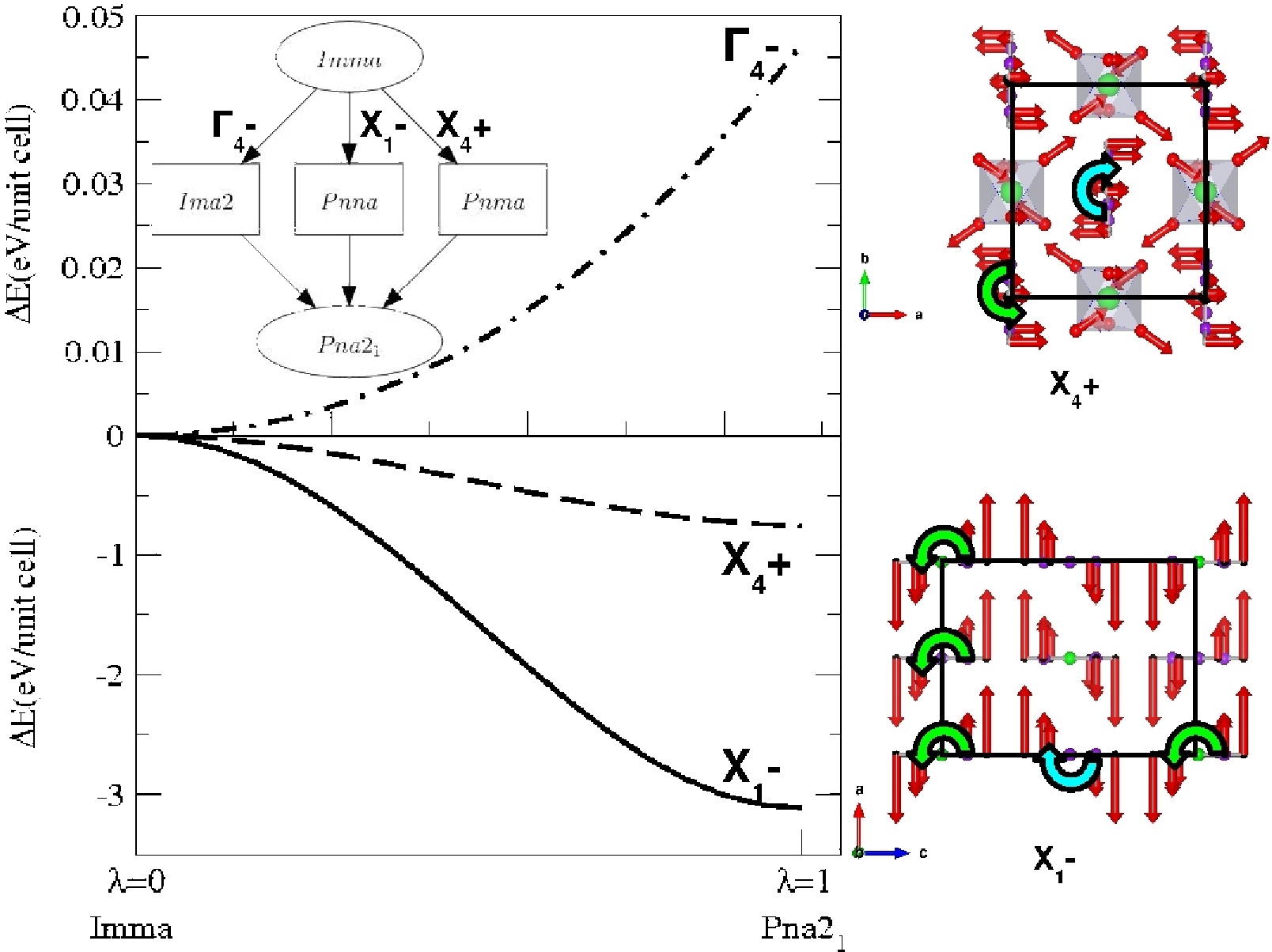}
\caption{ 
Variation of the total energy as a function 
of the different distortion modes. In the inset, 
the group-subgroup tree is shown, 
having the Pna2$_{1}$ as common maximal subgroup.
In the right part of the figure, projected view of the 
the distortion modes associated to non-polar instabilities. The length
of the arrows is proportional to the atomic displacements with respect 
to the Imma reference structure. The curved arrows denote clockwise or 
 counterclockwise  rotations.}\label{Fig3}
\end{figure}

Within the distortion mode connecting the Pnna structure 
to the Pna21 structure,  there can be atomic displacements hybridized 
with   true polar modes   which do not contribute to the electric 
polarization,
as it happens here with the JT distortion. In order to check this, 
  it is useful to introduce a higher supergroup for describing 
possible distortion modes and separate their role in the polar behavior.  In our case,
 the Pnna is pseudosymmetric with 
respect to a higher symmetry Imma space group. 
It is therefore possible to 
describe the polar structure with respect to this centric  Imma structure.
The global distortion relating the Imma and the
 Pna21 can be decomposed into three distinct atomic distortions: 
two zone-boundary modes at the $X$ point, transforming as 
the irreducible representions $X_{1}-$, $X_{4}+$, 
and a polar zone-center mode transforming as  
$\Gamma_{4}-$. They lower the symmetry to Pnna, Pnma and Ima2 
respectively (isotropy subgroups). Therefore, the first two modes do not produce 
any polarization  since the Pnna and
 Pnma are non-polar space group, while the only one 
producing the actual polarization is the  $\Gamma_{4-}$ mode. 
It is worth noting here 
that it is possible  to reach the Ima2 polar group without the zone-center 
polar instability, \textit{i.e.} only with 
a combined distortion  $X_{1}-$$\oplus$$X_{4}+$.\cite{rot2}
The relative size of symmetry-adapted mode amplitudes offers  
valuable clues to interpret the mechanism giving rise to the
electric polarization. The amplitude of these  modes $Q$ is very 
different: 0.57, 2.87 and 0.16 \AA\ for $X_{1}-$, $X_{4}+$ 
and $\Gamma_{4}-$ respectively. 
The fact that $Q_{X_{1}-}$ and $Q_{X_{4}+}$ are much larger than
$Q_{\Gamma_{4}-}$ suggests that the two first distortion modes are the
primary structural distortions with respect to the prototype
phase, acting as order parameters, while the polar mode is a
secondary induced distortion. To confirm this, we have calculated the variation
of the total energy with respect to the Imma structure 
(taken as   zero energy reference) 
as a function of the amplitude of each individual mode,
 Q$_{X_{1}-}$, Q$_{X_{4}+}$, and  Q$_{\Gamma_{4}-}$, 
as shown in Fig.\ \ref{Fig3}.

It is clear that $\Gamma_{4}-$ mode is stable and very soft,
 while the $X_{1}-$ and $X_{4}+$ are  unstable.
 In the same figure, inset from the left,  we also 
report the subgroup tree relation, 
which shows the chain of maximal subgroups connecting the Imma 
 and Pna2$_{1}$ space groups. 
 The latter appear 
as a common maximal subgroup of any pair of the space groups. 
It can be seen that the intersection of  
two mentioned isotropy subgroups is the observed Pna2$_{1}$  
space group, as it is the largest common subgroup of the  Pnna and
 Pnma space groups. This implies
 that the combination of these 
two non-polar modes is able to induce a polar 
phase (although they do not produce a polarization by themselves).

It is interesting to visualize the distortion modes, when viewed as atomic 
displacements with respect to the reference Imma structure. In the right 
part of Fig.\ \ref{Fig3}, we show only the $X_{1}-$ and $X_{4}+$ mode. The former 
shows significant displacements on the A-group molecule, and it represents 
a rotation around the $c$ axis which is anticlockwise  along the 
$a$ axis, and alternating clockwise (counterclockwise) along $c$ axis. The latter 
clearly represents the characteristic pattern of distortion of 
the $Q_{2}$ JT modes, 
which induces by hydrogen bondings\cite{Stroppa} 
other distortions on the A-group molecule, in form of 
a clock-wise rotation around  the $c$ axis. The $\Gamma_{4}-$ polar mode 
is not shown for simplicity, but it acts significantly 
on the A-group atoms.\cite{Stroppa} Further details about the mode visualization 
can be found in  the Supplementary  Information.

The presence of the polar distortion with a non-negligible significant 
amplitude can be explained as an induced effect through  symmetry-allowed anharmonic trilinear coupling with the primary non-polar 
distortion modes described as Q$_{X_{1-}}$Q$_{X_{4+}}$Q$_{\Gamma_{4-}}$. This simple coupling is sufficient to explain the presence of a non-zero amplitude 
Q$_{\Gamma_{4}-}$$\propto$Q$_{X_{1}-}$Q$_{X_{4}+}$ in order to minimize 
the energy, despite the mode itself being essentially stable. 
The combined $X_{4}+$ (JT distortion) $\oplus$ $X_{1}-$ (rotation of group A) is the 
``hybrid'' distortion $Q_{X_{41}}$ and it is what drives the 
system into the polar state. This means that  the polarization
 becomes non-zero only when both the guanidinium rotation  
  and JT distortions
condense. Examples of hybrid improper ferroelectrics
 have been recently found in inorganic material chemistry\cite{rot2}. 
However, they show more complicated 
topology than simple ABX$_{3}$ type 
and involve rotational modes such as tilting 
and rotations of octahedra.  Here, for the fist time,  
the hybrid improper ferroelectricity is found in a 
metal-organic framework in 
a simple ABX$_{3}$, and it involves a   
 Jahn-Teller mode  and rotational mode acting on the A-site.
It is important to highlight that the interaction between ferroelectric 
distortion and two rotational 
modes represent an important strategy for strong magnetoelectric coupling,
possibly at room temperature, as recently shown for inorganic transition-metal 
oxides.\cite{Ghosez} In MOF perovskite there is the additional advantages of the 
great tunability offered by the organic-inorganic duality, and the possibility 
of considering lead-free compounds, which is certainly an important aspect for 
environmental concerns.

We have shown in Fig. \ref{Fig2} (d)  that there is a  correlation between WFM component and 
the  ferroelectric polarization,  and therefore with the  JT
 distortions. In fact, i) a rotation of the AFD pattern switches the WFM  
from $+M (+P)$ to $-M(-P)$;
  ${\bf M}=0$ in the centrosymmetric Pnna reference structure,
 where the JT-induced antiferrodistortive pattern is zero but  the  octahedra remain 
strongly tilted.
This suggests that
 Dzyaloshinskii-Moriya interaction (DMI)
${\bf D}\cdot {\bf S}_i\times {\bf S}_j$,\cite{dzyaloshinskii,moriya} 
usually related to octahedra tilting,
may not be the source of WFM in this compound.
In fact, due to the AFM-A magnetic ordering,  
only DMI between spins belonging to different $ab$ planes is relevant;
each octahedron is tilted   from $c$ axis by an angle $\alpha$ $\sim$
31$^{\circ}$ , around $a$ axis: this implies  that 
the Dzyaloshinskii vector ${\bf D}_c$ is parallel to axis $a$,
as shown in inset from the left of  Fig. \ref{Fig2} (d).
By using symmetry analysis 
(see Supplementary Information)
it can be shown that only magnetoelectric trilinear coupling terms $P_c M_c L_a$ or $P_c M_a L_c$ are allowed in Cr-MOF, where $\bf P,M,L$ are
ferroelectric, ferromagnetic and antiferromagnetic order parameters respectively;
in both cases, the
cross product ${\bf S}_i\times {\bf S}_j$ is 
parallel to axis $b$ and perpendicular to
${\bf D}_c$, as shown in inset from the right of 
Fig. \ref{Fig2} (d).
Therefore ${\bf D}\cdot {\bf S}_i\times {\bf S}_j$ is zero,  
thus ruling out DMI as the source for the canted
antiferromagnetism. 

Our {\it ab initio} study  suggests a relevant role of JT distortions in combination with  spin-orbit coupling (SOC) interaction, for the origin of the canted antiferromagnetism.
Indeed, both JT and SOC involve the angular moment of the
electronic states\cite{bersuker}; furthermore, SOC is usually responsible for magnetic single-ion anisotropy (MSIA),
which has been devised by Moriya as an alternative origin for canted antiferromagnetism when
neighboring magnetic sites display different anisotropy axes\cite{moriya2}.
A second-order perturbation calculation
 (see Supplementary Information) shows that MSIA energy has the form $H_{msia}\,=\,E\,\sum_i\,
\left[ ({\bf S}_i\cdot {\bf s}_i)^2 - ({\bf S}_i\cdot{\bf l}_i)^2\right] + D\,\sum_i\,({\bf S}_i\cdot{\bf m}_i)^2$, where
${\bf l}_i,{\bf s}_i$ are unit vectors along the long and short bonds in M-octahedron basal plane and ${\bf m}_i$
a unit vector parallel to the M-O$_{ap}$ bond 
(see Fig. \ref{fig:mag}c), where  $E,D$  are the effective
magnetic anisotropy constants.
From this expression the interplay between SOC and JT interaction is readily seen, 
since the MSIA energy explicitly depends on the MO$_6$ octahedral JT distortions.
Therefore the local MSIA contribution is affected by the antiferrodistortive pattern of JT distortions and by the tilted
structure of the metal framework, both distortions
modifying the direction of anisotropies axes on each octahedron 
(see Supplementary Information).
Assuming classical spins,  which is a reasonable approximation for Cr spins,  
and considering MSIA as a small perturbation on the AFM-A magnetic configuration induced by symmetric exchanges $J_c, J_{ab}$,  the WFM can be described  by 
a small angle $\epsilon$ representing the  
 spin canting from the collinear spin
axis, \textit{i.e.} 
$\vert {\bf M}\vert = S \sin\epsilon$. By minimizing the mean-field energy, the
angle $\epsilon$ satisfies equation (see Supplementary Information):
\begin{eqnarray}\label{wfm_eps}
\tan2\epsilon= -\frac{2E\sin\alpha}{2J_c\mp D\cos^2\alpha},
\end{eqnarray}
where $\alpha$ is the tilting angle of octahedra and the sign $\mp$ is found for
 $L_a, L_c$ realizations of the AFM-A ordering respectively.
Eq. (\ref{wfm_eps}) shows that WFM appears only
when tilting distortions ($\alpha\neq0$) and JT-induced MSIA, through constant $E\neq 0$, are simultaneously present.
The WFM component points along $c$ ($a$) direction for $L_a$ ($L_c$) respectively, and it is reversed when
the antiferrodistortive pattern is rotated by $90^{\circ}$ 
with fixed tilting angle ($\alpha E\mapsto -\alpha E$),
in agreement with {\it ab-initio} results.

Since $D< J_c$, WFM is present irrespectively of the ratio $E/D$ provided
$E$ is nonzero: this explains the MSIA as the origin of 
the WFM component.
Note that the proposed mechanism is compatible with 
the trilinear coupling previously discussed, putting forward the
Jahn-Teller interaction as the main responsible for the coupling
 between $P$ and $M$
and for the predicted ME effect. The classical mean-field approximation
predicts that the $L_a,M_c$ realization of the canted AFM-A state
is energetically more stable
than the $L_c,M_a$ configuration, in agreement with DFT results, \textit{i.e.} the WFM points along the $c$ axis.
The MSIA parameters have been estimated from DFT calculations as $E\simeq 0.745\ meV$ and $D\simeq 0.113\
meV$; with these values a set of Monte Carlo calculations has been performed, confirming the mean-field analysis and
predicting a
critical temperature for the onset of the canted AFM-A at $T_c\simeq 40 K$ (see Fig.
\ref{fig:mag}e).

By using state-of-the-art-ab-initio calculations,
 we theoretically predict that
  the [C(NH$_{2}$)$_{3}$]Cr(HCOO)$_{3}$ MOF should be a new multiferroic 
 with a magnetoelectric coupling. For the first time in the MOF class
of compounds we show a peculiar origin of ferroelectricity, as due to a trilinear
coupling of two primary unstable non-polar modes with a secondary  stable 
and polar mode. Moreover, we highlight the important role of the Jahn-Teller 
antiferrodistortions in both the magnetoelectric and ferroelectric coupling.   
To support our interpretation we have 
carried out a symmetry mode analysis as well as 
we propose a microscopic model based on a perturbative approach.
These results    support the idea  that MOFs show 
promising new routes for achieving  multiferroic properties, an otherwise rare 
phenomenon in pure inorganic materials. Furthermore, a high tunability  of the 
ferroelectric polarization by means of the modification of the organic A cation has been recently shown in Ref. \cite{TuningJACS}.

{\bf Methods}\\
The calculations have been done by using the VASP package\cite{vasp,paw} 
with the Perdew-Burke-Ernzerhof
(PBE) GGA functional\cite{pbe}.
The energy cutoff was set to 400 eV and a 2$\times$4$\times$4 Monkhorst-Pack
grid of $k$-points was used. 
The Berry phase approach\cite{berry1}\
was employed  to calculate the ferroelectric polarization $P$.
Test calculations using the Heyd-Scuseria-Ernzerhof
hybrid functional (HSE)\cite{hse1} and with the Grimme 
corrections for weak-interactions\cite{Grimme}
have been considered:  the results are robust with 
respect to the different computational methods
giving us confidence in the reliability of the 
underlying  physical mechanisms.
Monte Carlo calculations were performed via
 standard Metropolis algorithm on a $28\times28\times28$ cubic
lattice with periodic boundary conditions. 
Symmetry analysis has been performed 
using the Bilbao Crystallographic Server\cite{BILBAO}, 
in particular using the 
symmetry software PSEUDOSYMMETRY\cite{PSEUDO}
 and AMPLIMODES\cite{AMPLIMODES1,AMPLIMODES2}

{\bf Acknowledgements}\\
This work has been supported by the 
European Community's Seventh Framework Programme FP7/2007-2013 under
grant agreement No. 203523-BISMUTH. A.S. wishes to thank Prof. I. B. Bersuker 
for useful insights about JT effect. A.S.  acknowledges
discussions with Dr. C. Autieri and Prof. E. Pavarini.
We acknowledge the CINECA award under the ISCRA initiative, for the
availability of high performance computing resources and support.
We gratefully acknowledge the very valuable help of Emre Tasci when using the
tools of the Bilbao Crystallographic Server for the mode analysis and its
visualization.
\begin{flushright}
Received: ((will be filled in by the editorial staff))\\
Revised: ((will be filled in by the editorial staff))\\
Published online: ((will be filled in by the editorial staff))
\end{flushright}

\section*{}
\clearpage

\begin{thebibliography}{80}

\bibitem{Yaghi} O. M. Yaghi, M. O'Keeffe, N. Ockwig, 
H. K. Chae, M. Eddaoudi, J. Kim,
\textit{Nature} \textbf{2003}, 423, 705.  

 

 
 
\bibitem{Ferey}  G\'{a}erard F\'{a}erey, A.K. Cheetham, \textit{Science}  \textbf{1999}, 283, 1125.
 
\bibitem{Ferey1} C. N. R. Rao, A. K. Cheetham, A. Thirumurugan,
 \textit{J. Phys.: Condens. Matter} \textbf{2008}, 20, 083202.

\bibitem{sumida} K. Sumida, D. L. Rogow, J. A.  Mason, T. M. McDonald, E. D.  Bloch, Z. R. Herm,
T.-H Bae, J. R. Long, Chem. Rev. \textbf{2012}, 112, 724.


 
\bibitem{Highlights}  M. J. Rosseinsky, \textit{Nat. Mat.} \textbf{2010}, 9, 609;

M. A. Green \textit{Nat. Mat.} \textbf{2010}, 9, 539;
2009 Metal-organic framewoks issue, Chem. Soc. Rev. \textbf{2009} 38, 1201.
 
\bibitem{Highlights1} H. Sato, R. Matsuda, K. Sugimoto, M. Takata, S.  Kitagawa, 
Nature Mater. \textbf{2010} 9, 661.

\bibitem{Highlights2} R. Makiura, S. Motoyama, Y. Umemura, H. Yamanaka, O. Sakata, H. Kitagawa, Nature Materials \textbf{2010} 9 565.

\bibitem{Highlights3} 
T. C. Narayan, T. Miyakai, S. Seki, M. Dinca,  
J. Am. Chem. Soc. \textbf{2012}, 31, 12932.


\bibitem{PRLMOF}
A. U. Ortiz, A. Boutin, A. H. Fuchs, F.-X. Coudert,
Phys. Rev. Lett. \textbf{2012}, 109, 195502.

\bibitem{Baburin} B. Assfour, S. Leoni, G. Seigert and I. A. Baburin, 
Adv. Mat. \textbf{2011}, 23, 1237.

\bibitem{Tsukerblat}
C. Bosch-Serrano, J.M. Clemente-Juan, E. Coronado, A. Gaita-Arino, A. Palii, B.  Tsukerblat,
Phys. Rev. B \textbf{86} 024432 (2012).

\bibitem{NewRoutes} 
R. Ramesh, \textit{Nature}, \textbf{2009}, 461 , 1218.

\bibitem{NewRoutes1} 
T. Besara, P. Jain, N. S. Dalal, P. L. Kuhns, A. P. Reyes, H. W. Kroto, A. K. Cheetham, Proc. Natl. Acad. Sci. U. S. A. \textbf{2011} 108, 6828.

\bibitem{NewRoutes2} P. Jain, V. Ramachandran, R.J. Clark, H.D. Zhou, B.H. Toby, N.S. Dalal, H.W. Kroto, A.K. Cheetham, J. Am. Chem. Soc. \textbf{2009}, 131, 13625.


\bibitem{NewRoutes3}  X. Guan-Cheng, Z. Wen, M. Xiao-Ming, C. Yi-Hong, Z. Li, C. Hong-Ling, W. Zhe-Ming, X. Ren-Gen, G. Song  J. Am. Chem. Soc., \textbf{2011}, 133 14948. 

\bibitem{Spaldin} N. A. Spaldin, M. Fiebig, \textit{Science} \textbf{2005}
309, 391.


\bibitem{cheongmostovoy} 

S.-W.~Cheong, M.~Mostovoy, \textit{Nat. Mater.} \textbf{2007}, 6, 13.


\bibitem{BFO}  J. Wang, J. B. Neaton, H. Zheng, V. Nagarajan, S. B.
 Ogale, B. Liu, D. Viehland, V. Vaithyanathan, D. G.
 Schlom, U. V. Waghmare, N. A. Spaldin, K. M. Rabe, M. Wuttig, 
 R. Ramesh, 
\textit{Science} \textbf{2003}, 299, 1719.
 
\bibitem{rot1}  E. Bousquet, M. Dawber, N. Stucki, C. Lichtensteiger,
 P. Hermet, S. Gariglio, J.-M. Triscone, Ph. Ghosez
\textit{Nature (London)} \textbf{2008}, 452, 732.

 
\bibitem{rot2}  
N. A. Benedek  and C. J. Fennie, 
Phys. Rev. Lett. \textbf{2011}, 106, 107204.

\bibitem{rotexp} C. Adamo, private communication.

\bibitem{rot3} 
J. Lopez-Perez, J. Iniguez, 
Phys. Rev. B \textbf{2011}, 84, 075121.
 

\bibitem{rot4} J. M. Rondinelli, C. J. Fennie, Adv. Mat. \textbf{2012} 24,
1961.

\bibitem{rot5} T. Fukushima, A. Stroppa, S. Picozzi, J. M. Perez-Mato, 
Phys. Chem. Chem. Phys. \textbf{2011}, 13, 12186.


\bibitem{KeLi} K.-L. Hu, M. Kurmoo, M. Wang, S. Gao, 
\textit{Chem. Eur.} \textbf{2009}, 15, 12050.


\bibitem{Stroppa} A. Stroppa, P. Jain, P. Barone, M. Marsman, J.M. Perez-
Mato, A. K. Cheetham, H. W. Kroto, S. Picozzi, Angew.
Chem. Int. Ed. \textbf{2011}, 123, 5969.

\bibitem{moriya2} T. Moriya, Phys. Rev. \textbf{1969}, 120 , 91.

\bibitem{Tc1} L.-S. Long, X.-M. Chen, M.-L. Tong, Z.-G. Sun,
Y.-P. Ren, R.-B. Huang, L.-S. Zheng,  
J. Chem. Soc., Dalton Trans. \textbf{2001}, 2888.


\bibitem{Tc2} 
M. Kurmoo, H. Kumagai, 
S. M. Hughes and C. J. Kepert, 
Inorg. Chem.  \textbf{2003}, 42, 6709. 

\bibitem{Margadonna} 
S. Margadonna, G.  Karotsis, 
J. Am. Chem. Soc. \textbf{2006}, 128, 16436.
 

\bibitem{phi1} G. Matsumoto, J. Phys. Soc. Jpn. \textbf{1970}, 29, 606.

\bibitem{phi2} J. Kanamori, J. Appl. Phys. \textbf{1960}, 31, 14S.

\bibitem{Lines} M. E. Lines and A. M. Lines, 
{\it Principles and applications of ferroelectrics and related materials},
Clarendon. Press, Oxford \textbf{1977}. 

\bibitem{bersuker}  I. Bersuker,  {\it The Jahn-Teller Effect}, Cambridge University Press, Cambridge, \textbf{2006}.

\bibitem{GKA} J. B. Goodenough,   
\textit{Magnetism and Chemical Bond}, Interscience Publ., N.Y.-Lnd., \textbf{1963};
~D.~J.  Khomskii,   
in \textit{Spin Electronics}
(Eds: M. Ziese and M.J. Thornton), Springer-Verlag, Berlin Heidelberg \textbf{2001}.


\bibitem{towler1995} ~M. Towler, ~R. Dovesi, ~V.~S. Saunders,
\emph{Phys. Rev. B} \textbf{1995}, 52, 10150.

\bibitem{Ghosez} P. Ghosez, J.M. Triscone, Nat. Mat. \textbf{2011}, 10, 269.


\bibitem{dzyaloshinskii} I. Dzyaloshinsky, 
J. Phys. Chem. Sol. \textbf{1958}, 4, 241 .

\bibitem{moriya} 
T. Moriya, Phys. Rev. Lett \textbf{1960}, 4, 228.

 
\bibitem{vasp}  G. Kresse and J. Furthm\"{u}ller, 
Phys. Rev. B \textbf{1996}, 54, 11169.

\bibitem{paw} P. E. Bl\"{o}chl, Phys. Rev. B \textbf{1994}, 50, 17953.

\bibitem{pbe} J. P. Perdew, K. Burke, M. Ernzerhof, 
Phys. Rev. Lett. \textbf{1996}, 77, 3865.

\bibitem{berry1} R. D. King-Smith, 
D. Vanderbilt, Phys. Rev. B \textbf{1993}, 47, 1651.


\bibitem{hse1} J. Heid, G. E. Scuseria, M. Ernzerhof, 
J. Chem. Phys. \textbf{2003}, 118, 8207.

\bibitem{Grimme} S. Grimme, J. Comput. Chem. {\bf 2004}, 25, 1463.

\bibitem{BILBAO} http://www.cryst.ehu.es/

\bibitem{PSEUDO} 
C. Capillas, E.S. Tasci, G. de la Flor, D. Orobengoa, J.M. Perez-Mato and M.I. Aroyo,
 \textbf{2011} Z. Krist., 226, 186.

\bibitem{AMPLIMODES1} D. Orobengoa, C. Capillas, M. I. Aroyo and 
 J.M. Perez-Mato \textbf{2009} J. Appl. Cryst.  A42, 820.


\bibitem{AMPLIMODES2} J.M. Perez-Mato, D. Orobengoa and M.I. Aroyo,
Acta Cryst A \textbf{2010} 66 558.

\bibitem{TuningJACS} D. Di Sante, A. Stroppa, P. Jain, S. Picozzi, 
J. Am. Chem. Soc. \textbf{2013}  135  18126.
\end{thebibliography}
\end{document}